# Easy and Fast Design and Implementation of PostgreySQL based image handling application

[1]Kisor Ray  [2]Sourav Bag  [3]Saumen Sarkar
*Techno India Agartala,Tripura University   Techno India Agartala,Tripura University   NSEC (Techno India Group),WBUT*

*Abstract—* *In modern computing, RDBMS are great to store different types of data. To a developer, one of the major objectives is to provide a very low cost and easy to use solution to an existing problem. While commercial databases are more easy to use along with their new as well as documented features come with complicated licensing cost, free open source databases are not that straightforward under many situations. This paper shows how a completely free advanced open source RDBMS like PostgreSQL could be designed and modified to store and retrieve high quality images in order to use them along with a frontend application.*

## I. INTRODUCTION

Storing of images in a database efficiently and then retrieve and use them over a network is always a challenging task.This challenge becomes harder should the database belong to an open source RDBMS category and available completely free. Though a relatively easy solution is to use a commercial RDBMS for which cost could become prohibitive especially for a budding institute.In this paper we will examine how we can create a frontend application very easily using an advanced open source RDBMS like PostgreSQL where images could be stored , retrieved easily with a purpose to use them with the frontend multiuser application in a network environment.

## II. MOTIVATION

In our institute, when we decided to build up an application which can take care of many of our routine tasks, the student & workforce module were harder than our initial anticipation because we had decided to store the images of our students and staff members along with their data in the backend database. For the purpose of rapid development and easy to use report generation, we had opted for VBA for the frontend application development. After a careful initial research, we then opted for PostgreSQL 9.4 as our backend database. While developing our application and specially storing of images in the backend postgresql, we adopted certain mechanisms which helped us to develop the application very fast . We thought many budding institutes similar to ours may require the similar type of application and may adopt the same technique to develop their custom application without any additional cost.

## III. BACKGROUND

For our institute we needed a number of modules for our routine business. Among these the student and workforce modules needed to store data and individual images with an objective to store every important information which could be used for many operations including but not limited to contact, score, personal behaviour, leave, fees, barcoded photo identity cards, HR etc.. We have used PostgreSQL 9.4 initially on a Windows 7 environment along with VBA (MS access 2010) at the development stage and then PostgreSQL 9.4 under Linux (Ubuntu 12.04 LTS) as a backend database server and compiled version of frontend (MDE) at individual workstations in an wi-fi network environment. Each workstation needs appropriate PostgreySQL ODBC driver (psqlodbc 32 or 64 bit) installed into it. The initial database design was done with the MSAccess 2010. The tables were then migrated to the backend PostgreSQL database manually. While developing our frontend forms, we have used one ActiveX component named as AccessImagine to handle the images at the form and report level. The image field at the frontend was taken as OLE Object which was mapped to the backend postgresql's custom lo type field.

## IV. DESIGN CONCEPT

One of the main important point in this paper is that how the tables using PostgreSQL was designed to handle the images efficiently and how it was implemented along with the vba frontend with activex component ? Object identifiers (OIDs) are used internally by PostgreSQL as primary keys for various system tables. OIDs are not added to user-created tables, unless WITH OIDS is specified when the table is created, or the default_with_oids configuration variable is enabled. Type oid represents an object identifier [1]. If the field uses a PostgreSQL domain (rather than a basic type), it is the OID of the domain's underlying type that is returned, rather than the OID of the domain itself[2]. A domain is essentially a data type with optional constraints. So, should we decide to store the images in the backend PostgreySQL through our OLE Object type captured at the frontend, we need to define a large object type (lo) .However, we need to create a domain of the object type as OID before we add a custom column type where we intend to store our images.

For example, we have a database named as TIA02 which is created under the postgraysql.TIA02 has two specific tables Tab_T_ Emp_Details and Tab_T_Students where we need to store the images in a field xxPhoto. We then make and execute the following scripts :

   CREATE DOMAIN lo as oid;

   ALTER TABLE "Tab_T_Emp_Details"
   ADD COLUMN myphoto  lo;

   ALTER TABLE " Tab_T_Students "
   ADD COLUMN myphoto  lo;

It may please be noted that the creation of the 'domain lo as oid' is only needed once at the top level of the schema, then 'lo' type filed could be added to any table under that schema as column.

Fig 1: Execution of query on PostgreySQL using PgAdminIII Query Tool

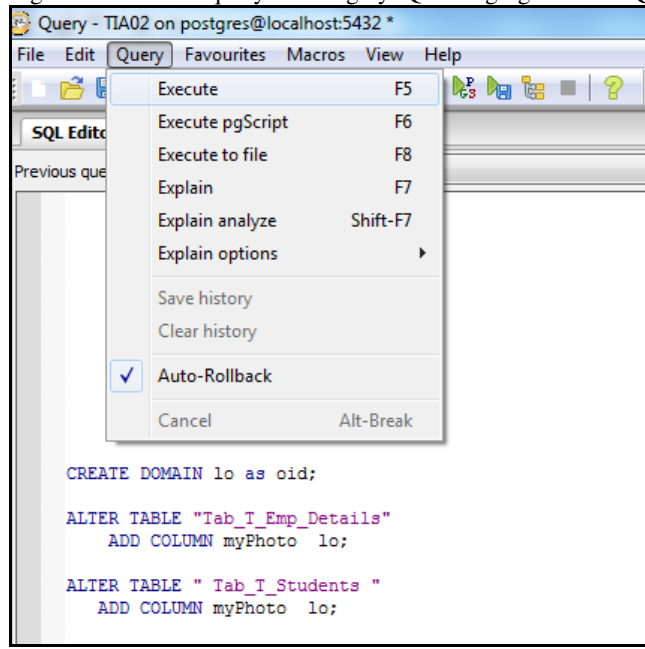

V. IMPLEMENTATION

Brief structures of the two tables for the employee and the students are given below. The employee/student data is entered through the frontend application's form which uses the ActiveX component AccessImagine to capture and display the images on the form and reports.The employee and student tables are having a OLE Object fields named as Emp_Photo and Stu_Photo respectively.The VBA frontend connects the backend PostgreySQL database as linked table.The image fields at the the frontend are mapped to the backend PostgreySQL tables. The OLE Object type filed is mapped to the custom 'lo' types at the backend.

TABLE I
Database Table Structure : Tab_T_EmpDetails

| Field Name | Data Type | Field Size | Remark |
|---|---|---|---|
| Emp_ID | Number | Long Integer | |
| First_Name | Text | 50 | |
| Middle_Name | Text | 50 | |
| Last_Name | Text | 50 | |
| Emp_Contact_No | Number | Double | |
| Date_Of_Birth | Date | | |
| Dept | Text | 20 | |
| Date_Of_Joining | Date | | |
| Highest_Education | Text | 20 | |
| Designation | Text | 20 | |
| Employment_Type | Text | 20 | |
| Gender | Text | 6 | |
| Blood_Group | Text | 6 | |
| Years_of_Experience | Number | Integer | |
| myphoto | OLE Object | | Mapped to PostGreySQL LO type |
| Remark | Text | 255 | |

TABLE II
Database Table Structure : Tab_T_Students

| Field Name | Data Type | Field Size | Remark |
|---|---|---|---|
| Student_ID | Number | Long Integer | |
| First_Name | Text | 50 | |
| Middle_Name | Text | 50 | |
| Last_Name | Text | 50 | |
| Student_Contact_No | Number | Double | |
| Date_Of_Birth | Date | | |
| Branch | Text | 20 | |
| Date_Of_Admission | Date | | |
| Session | Text | 20 | |
| Semester | Number | Integer | |
| AICTE_Course_ID | Number | Long Integer | |
| Gender | Text | 6 | |
| Blood_Group | Text | 6 | |
| myphoto | OLE Object | | Mapped to PostGreySQL LO type |
| Remark | Text | 255 | |

Fig 2 : Table Tab_T_EmpDetails in PostgreySQL

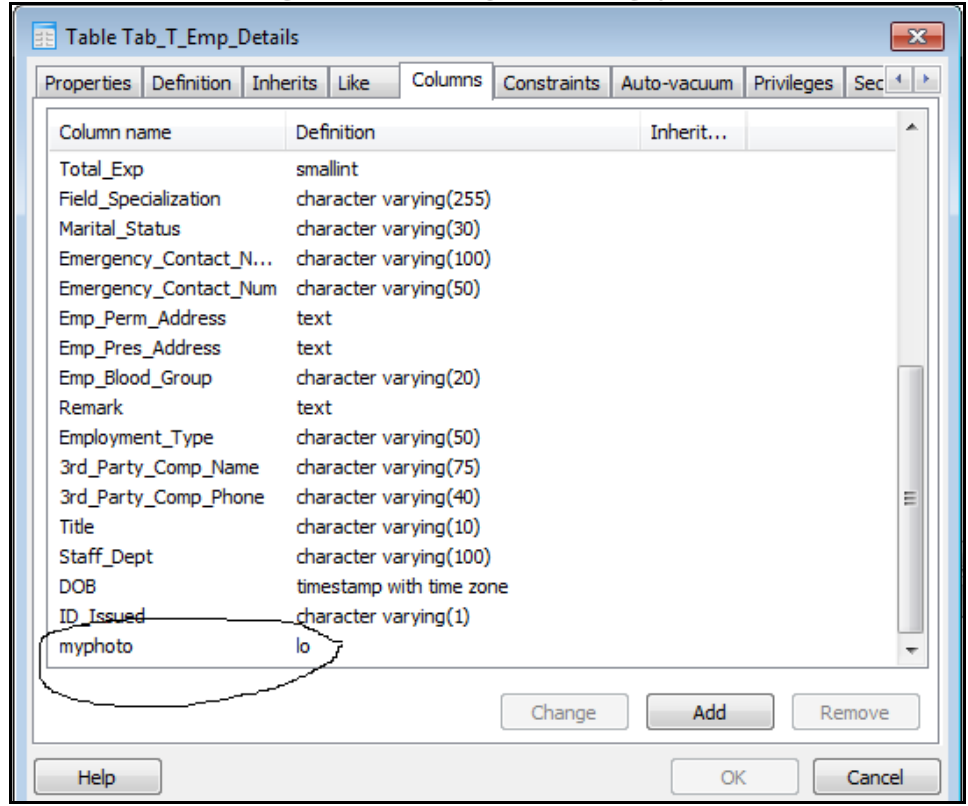

In the above figure , myphoto is a column of 'lo' type added to the EmpDetails table .

Data is entered into the PostgreSQL database using the forms at the VBA based frontend. This data could be edited and updated and images could be easily accommodated and modified using the auto corp feature.

Fig 3 : Employee Details Entry Form

Fig 3 shows the form at the frontend where all types of data including image could be inserted into the data base. This frontend also has the facility of directly capturing any image from the camera connected to the workstation through the usb port.

The data stored in the database could be easily used for different types of processing including generation of barcoded identity cards etc. The identity card uses the image from the backend PostgreSQL database and generates the barcode in graphics mode (only in print mode) using a separate barcode module. These cards are generated with the use of MS Access Report & VBA codes.

Fig 4 : Employee Identity Card

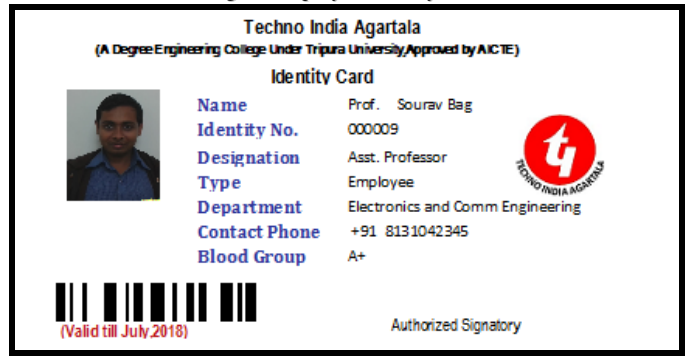

VI. CONCLUSION

While developing a cost effecting solution for the present and future use, we decided to go for PostgreySQL which is one of the most advanced and powerful open source free RDBMS. In the process of development, we have tried to implement number of techniques including declaring postgreyfield as 'bytea'. Since, we did not get the desired solution, we opted to create custom data type 'lo' using the domain and oid. That served our purpose along with the use of the ActiveX component AccessImagine at the frontend application. The process of design and development described in this paper provides the developer with the ability to build an application rapidly which not only serves the purpose but also quite reliable. Unlike MS Access, PostgreSQL provides a huge repository at free of cost for all types of data.

VII. **FURTHER WORK**

The IMAGE datatype can be added to a PostgreySQL database using the Postgres extension mechanism:
CREATE EXTENSION pg_image; CREATE EXTENSION pg_image;[3] .Many functions and operators are provided for working with values of type IMAGE. We did not explore this extension mechanism.